\documentclass{aa}
\usepackage{graphicx}
\begin{document}

\title{A high frequency radio study of G11.2$-$0.3, a historical
supernova remnant with a flat spectrum core}

\author{R. Kothes
          \inst{1}
          \and
          W. Reich\inst{2}}

\offprints{R. Kothes}

\institute{National Research Council, Herzberg Institute of Astrophysics,
              Dominion Radio Astrophysical Observatory,
              P.O. Box 248, Penticton, British Columbia, V2A 6K3, Canada\\
              email: roland.kothes@hia.nrc.ca
           \and
             Max-Planck-Institut f\"ur Radioastronomie, Auf dem H\"ugel 69,
             53121 Bonn, Germany\\
             email: wreich@mpifr-bonn.mpg.de}

\date{Received 30 January 2001 / Accepted 19 March 2001 }

\abstract{
We present radio maps of the historical supernova remnant
G11.2$-$0.3 in the frequency range from 4.85~GHz to 32~GHz. The
integrated spectrum with $\alpha = -0.50$ (S $\sim \nu^\alpha$) is
dominated by its steep spectrum shell emission ($\alpha \sim -0.57$),
although a flat spectrum core structure classifies G11.2$-$0.3 as a
composite supernova remnant. A radial magnetic field structure is
observed. An analysis of the multi--frequency polarization data results
in highly varying rotation measures along the shell. The percentage
polarization is rather low ($\sim$ 2~\%) and we conclude that
G11.2$-$0.3 is in the transient phase from free to adiabatic expansion.
The central flat spectrum component is partly resolved. A compact radio
source with an inverted spectrum likely coincides with the previously
detected X--ray pulsar (Torii et al.\ \cite{torii1997}). Two symmetric
structures with flat radio spectra possibly indicate a bipolar
outflow. Combining available X--ray and radio data we conclude that G11.2$-$0.3
is likely the remnant of a type II supernova explosion with an early
type B progenitor star.
 }

\titlerunning{A radio study of G11.2$-$0.3}

\maketitle
%

\section{Introduction}

Only a small number of young supernova remnants (SNRs) has been
identified in our Galaxy so far. One of them is the non--thermal source
G11.2$-$0.3. Clark \& Stephenson (\cite{clark}) associated this SNR
with the supernova explosion of {\scriptsize A.D.}~386 observed by
Chinese astronomers during the Chin Dynasty. In the radio continuum
G11.2$-$0.3 shows the typical structure of a shell--type SNR with a
spectral index of $\alpha \approx -0.5$ (S $\sim \nu^{\alpha}$). Green
et al. (\cite{green}) reported a clumpy shell and the lack of a sharp
outer boundary based on high resolution VLA data. This indicates a SNR
in an early evolutionary state. Green et al. (\cite{green}) found no
evidence for spectral index variations between 1.4~GHz and 5~GHz.
However, Morsi \& Reich (\cite{morsi}) reported on a flat spectrum core
based on early 32~GHz observations with the Effelsberg 100-m
radio telescope.

\ion{H}{i}--absorption measurements towards the SNR indicate a
distance of about 5~kpc (Radhakrishnan et al., \cite{radha}; Green
et al.\ \cite{green} after reinterpretation of the absorption
spectrum measured by Becker et al.\ \cite{becker1985}). At this
distance G11.2$-$0.3 has a diameter of approximately 6~pc, which
is comparable to other historical SNRs like Kepler and Cas~A with
$\rm D\approx 4$~pc and Tycho with $\rm D\approx 6$~pc.

\begin{figure}
\centering
\includegraphics[bb = 45 120 570 605,width=8.8cm,clip]{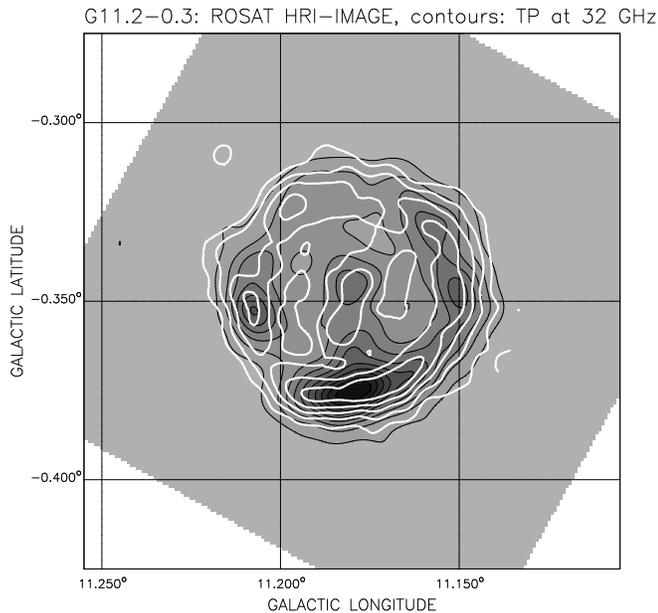}
\caption{White contours representing the Effelsberg 32~GHz
observation of the SNR G11.2$-$0.3 superimposed on a greyscale
image of the ROSAT HRI measurement. The image is shown with
Galactic coordinates and thus rotated from later images}
   \label{xrayradio}
\end{figure}

X--ray observations of G11.2$-$0.3 show a soft shell--like structure
which is most prominent in the southeast (Reynolds et al.\
\cite{reynolds}, see Fig. \ref{xrayradio}). In addition a
hard spectrum core exists (Vasisht et
al.\ \cite{vasisht}), which was interpreted as the synchrotron nebula
of an embedded pulsar. This pulsar was subsequently discovered by Torii
et al. (\cite{torii1997}) through observations with the {\it Advanced
Satellite for Cosmology and Astrophysics} (ASCA). It is a young pulsar
with a period of 65~ms. The rotational energy loss rate was calculated
to be $\dot{\rm E} = 8.8^{+9.9}_{-2.6} \times 10^{36}$~erg/s from
follow--up X--ray observations (Torii et al.\ \cite{torii1999}). This
is a rather low value for a young pulsar and indicates a low radio
surface brightness for the surrounding synchrotron nebula (Kothes\
\cite{kothes}).

In order to detail the properties of G11.2$-$0.3, in particular
its central component and its polarization structure, we conducted
a number of new observations at high radio frequencies.

\section{Observations}

The Effelsberg 100-m radio telescope was used to map fields around
G11.2$-$0.3 at high radio frequencies. In 1996 observations at 4.85~GHz
and 14.7~GHz were carried out as part of test measurements for newly
installed receiver systems. In 1999 observations at 10.45~GHz and 32~GHz
were added. At 32~GHz a new receiver system was used with improved
sensitivity compared to Morsi \& Reich's (\cite{morsi}) measurements
at the same frequency. All receiver systems used were installed in the
secondary focus of the telescope. The 4.85~GHz, 10.45~GHz, and 32~GHz
observations were made by recording both circularly polarized
components to obtain total intensity. IF correlation of the circular
components results in linear polarization data. At 14.7~GHz just total
intensity channels were available. The receiving systems consist of two
feeds at 4.85~GHz, four feeds at 10.45~GHz and 14.7~GHz, and three
feeds at 32~GHz. Observational parameters are listed in
Table~\ref{obspara}. All systems have highly stable cooled
HEMT receivers at their front--ends and run in total power mode. Data
reduction relies on application of the ``software beam
switching'' technique to reject influences of weather effects
(Morsi \& Reich\ \cite{morsi2}).

\begin{table}
  \caption{Observational parameters}
  \label{obspara}
  {
  \begin{tabular}{lllll}\hline
   & & & & \\
  Frequency [GHz] & 4.85 & 10.45 & 14.7 & 32\\
   & & & & \\ \hline
   & & & & \\
  Feeds & 2 & 4 & 4 & 3 \\
  T$_{\rm sys}$ [K] & 30 & 50 & 200 & 130\\
  Bandwidth [MHz] & 500 & 300 & 1000 & 2000 \\
  HPBW [\arcmin ] & 2.5 & 1.1 & 0.86 & 0.45 \\
  Velocity [\arcmin /min] & 60 & 40 & 25 & 20\\
  Step Interval [\arcsec ] & 60 & 20 & 15 & 12\\
  Mapsize [$\arcmin \times \arcmin$] & 30$\times$20 & 30$\times$10 & 12$\times$12
       & 15$\times$6.5 \\
  Observation Date & 2.96 & 8.99 & 1.96 & 4./5.99 \\
  Calibrator I & 3C286 & 3C286 & NGC7027 & 3C286\\
  Flux Density [Jy] & 7.5 & 4.5 & 6.2 & 2.1\\
  Calibrator PI & 3C286 & 3C286 & -- & 3C286 \\
  Linear Pol. [\%] & 11.5 & 11.7 & -- & 12.3 \\
  Pol. Angle [$\degr$] & 33 & 33 & -- & 33\\
  Coverages & 1 & 1 & 1 & 8 \\
  RMS-I [mJy/b.a.] & 19 & 6 & 10 & 7.5\\
  RMS-PI [mJy/b.a.] & 6 & 2.5 & -- & 2.0\\
   & & & & \\ \hline
  \end{tabular}}
\end{table}

All observations were made in the equatorial coordinate system (B1950).
The scan direction was along the azimuth. The standard data reduction software
package based on the NOD2 format has been applied. The multi--feed
observations at 14.7~GHz and 32~GHz were restored using the algorithm
described by Emerson et al. (\cite{emerson}). At 4.85~GHz and 10.45~GHz
data from different feeds were reduced separately and combined
afterwards. Baseline improvement by unsharp masking (Sofue \& Reich\
\cite{sofue}) was applied to all observations. The ``Plait''
algorithm described by Emerson \& Gr\"ave (\cite{emerson88}) was
used to combine the numerous coverages observed at 32~GHz at different
parallactic angles by destriping the maps in the Fourier plane. This
increases the signal--to--noise ratio of the final map substantially.

We made also use of an unpublished 23~GHz map with an angular
resolution of 1\farcm 2 observed in 1987 by W. Reich, P. Reich,
and Y. Sofue with the Nobeyama 45-m radiotelescope. A cooled HEMT
receiver was used and a map was made from two coverages at
different parallactic angles. It was calibrated relative to
NGC\,7027 assuming 5.7~Jy. The reduction of the double--beam
observations was described by P. Reich (\cite{pr}).

\begin{figure*}
\centering
\includegraphics[bb = 57 300 540 790,width=18cm,clip]{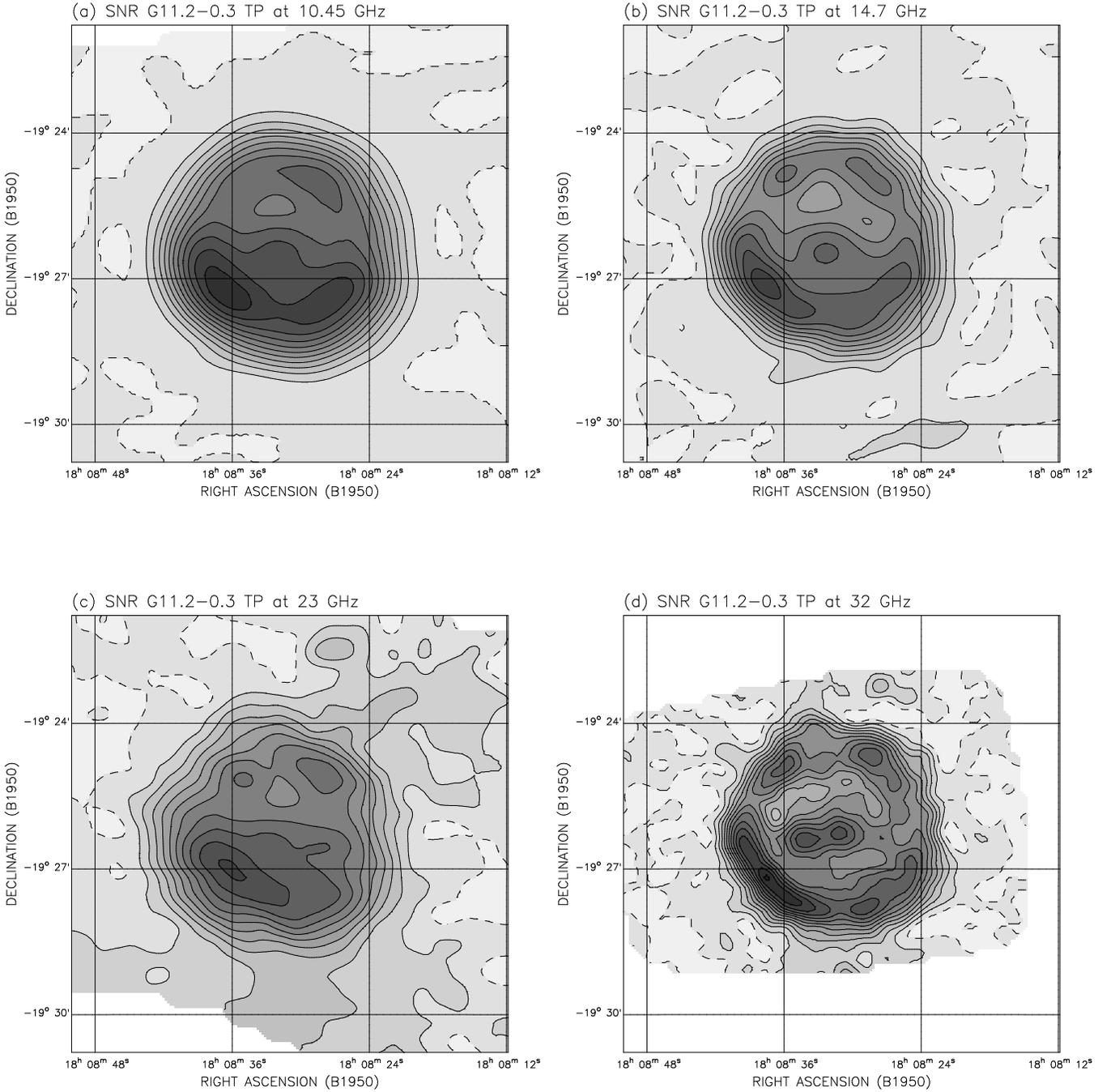}
\caption[]{SNR G11.2$-$0.3 in total intensity. The angular
resolutions of the maps are listed in Table \ref{obspara}.
Contours are at 0~mJy/beam (dashed contour), and at:
50--600~mJy/beam in steps of 50~mJy/beam (10.45~GHz and 23~GHz),
40--400~mJy/beam in steps of 40~mJy/beam (14.7~GHz),
10--120~mJy/beam in steps of 10~mJy/beam (32~GHz)}
  \label{tpmaps}
\end{figure*}

\begin{figure*}
\centering
\includegraphics[bb = 55 382 545 610,width=18cm,clip]{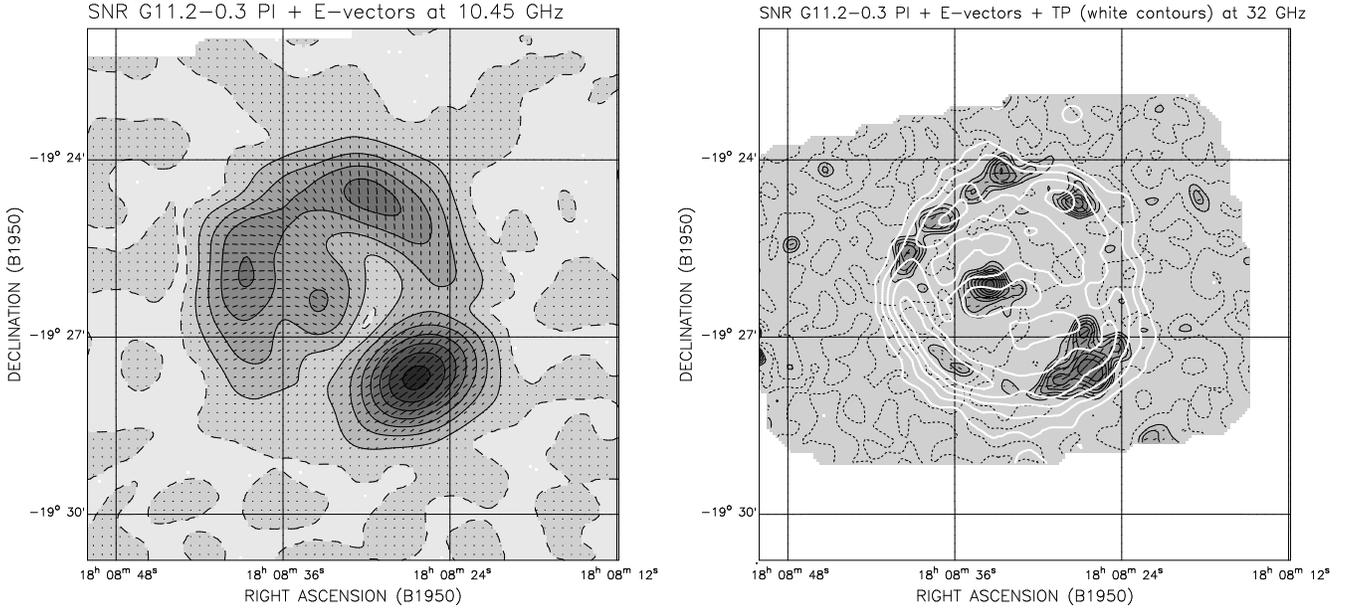}
\caption{Radio maps of G11.2$-$0.3 in polarized intensity with
vectors in E--field direction. Contours are at 0~mJy/beam (dashed
contour), and at 2.5--27.5~mJy/beam in steps of 2.5~mJy/beam
(10.45~GHz); 2--7.6~mJy/beam in steps of 0.8~mJy/beam (32~GHz).
The white contours represent the total power emission at 32~GHz}
   \label{pimaps}
\end{figure*}

\begin{figure*}
\centering
\includegraphics[bb = 57 320 530 545,width=18cm,clip]{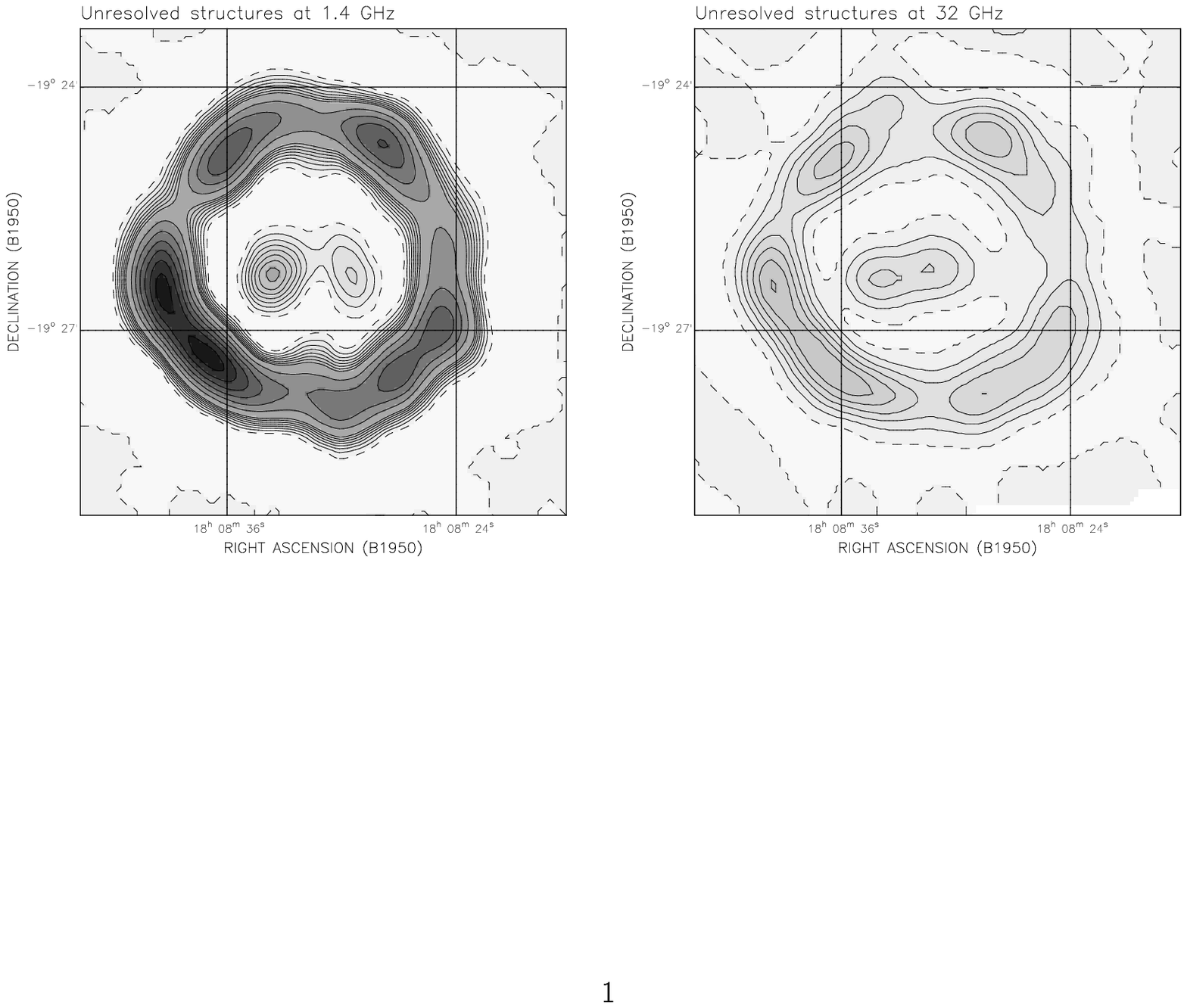}
\caption[]{Unresolved structures of the SNR G11.2$-$0.3 at
1.4~GHz, taken from the NVSS, and at 32~GHz. Diffuse structures
were removed following Sofue \& Reich (\cite{sofue}). Both maps
have a resolution of $45\arcsec$. The contour levels are at
0~mJy/beam (dashed line), at 10--70~mJy/beam in steps of
10~mJy/beam, and at 110--310~mJy/beam in steps of 40~mJy/beam }
   \label{interior}
\end{figure*}

\section{Results}

The total intensity maps at 10.45~GHz, 14.7~GHz, 23~GHz and 32~GHz
are shown in Fig.~\ref{tpmaps}. In Fig.~\ref{pimaps} polarized
intensities at 10.45~GHz and 32~GHz are displayed.

\subsection{The radio emission from the shell}

The overall structure of the radio continuum emission is most evident
in the 32~GHz map (Fig.~\ref{tpmaps}d), where the angular resolution is
highest. The SNR shows an almost circular shell consisting of four prominent
sectors, with the strongest towards the southeast. Three of
those sectors show clear polarization with electric field vectors
perpendicular to the expansion direction (Fig.~\ref{pimaps}).
The southeastern sector is almost unpolarized.

\subsection{The interior structure}

Within the shell there are two slightly extended sources visible at
32~GHz (see Fig.~\ref{tpmaps}d). On our lower frequency maps these
sources appear unresolved and confused with the shell as a result
of the lower angular resolution. Both inner sources seem to be polarized
(see Fig.~\ref{pimaps}, Table~3) but a more prominent polarization
feature is located slightly north of them. It has no distinct
counterpart in total intensity and seems to be related to diffuse
emission either from the shell or the inner structure of the SNR.

In order to measure the spectrum of the two compact sources
we used the NVSS data at 1.4~GHz (Condon et al.\ \cite{condon})
having an angular resolution of $45\arcsec$. We applied the ``background
filtering technique'' (Sofue \& Reich\ \cite{sofue}) to separate
the diffuse emission from the shell which is confused with the inner
structure. We applied the method to the NVSS map and the 32~GHz map,
convolved to $45\arcsec$, in exactly the same way to get comparable
results. The resulting maps are displayed in Fig.~\ref{interior}.

The NVSS map shows two clearly separated sources, while at 32~GHz
a bar-like structure with two emission peaks is visible. This
results from an additional component showing up at 32~GHz, which
is located between the two sources visible at 1.4~GHz. In the
observations of Green et al. (\cite{green}) there is a slightly
extended source visible at the very center of the 4.76 GHz map
which we believe has an inverse spectrum because it is not visible
in their 1.46 GHz map. There are only the two extended diffuse
structures which are also visible at 4.76~GHz.

\begin{figure}[htb]
\centering
\includegraphics[bb = 30 150 580 705,height=8cm,clip]{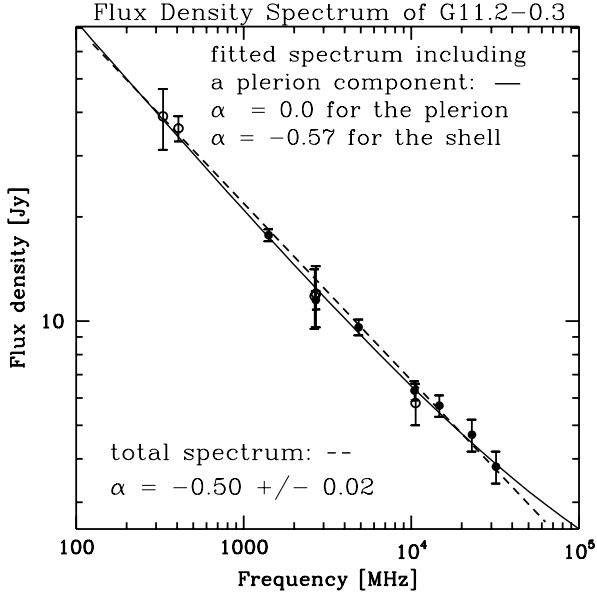}
\caption{The spectrum of G11.2$-$0.3. Integrated flux densities as
listed in Table~2 are shown by filled circles. Other values (open
circles) were taken from Kassim (\cite{kassim}), Goss \& Day
(\cite{goss}), Becker \& Kundu (\cite{becker1975}), Shaver \& Goss
(\cite{shaver}), and Milne et al. (\cite{milne}). The fits to the
data are discussed in Sect.~3.3}
   \label{spectrum}
\end{figure}

\begin{table}
\caption{Integrated total power and polarization flux densities.
The data at 1408~MHz and 2695~MHz are taken from the source lists of
the Effelsberg Galactic Plane surveys (Reich et al.\ \cite{reich1},
\cite{reich2}).}
  \label{flux}
  {
  \begin{tabular}{rr@{$\pm$}lr@{$\pm$}lr@{$\pm$}l}\hline
  \ Frequency & \multicolumn{2}{l} \bf TP & \multicolumn{2}{l}
  \ PI & \multicolumn{2}{l} \bf \%-Pol. \\
  \ [MHz] & \multicolumn{2}{l} \bf [Jy] & \multicolumn{2}{l}
  \ [mJy] & \multicolumn{2}{l} \bf \\ \hline
  1408 & 17.7 & 0.7 & \multicolumn{2}{c} -- & \multicolumn{2}{c} -- \\
  2695 & 11.5 & 0.5 & \multicolumn{2}{c} -- & \multicolumn{2}{c} -- \\
  4850 & 9.6 & 0.5 & 220 & 30 & 2.3 & 0.4 \\
  10450 & 6.3 & 0.4 & 140 & 10 & 2.2 & 0.3 \\
  14700 & 5.7 & 0.4 & \multicolumn{2}{c} -- & \multicolumn{2}{c} -- \\
  23000 & 4.7 & 0.5 & \multicolumn{2}{c} -- & \multicolumn{2}{c} -- \\
  32000 & 3.8 & 0.4 & 90 & 8 & 2.5 & 0.4 \\ \hline
  \end{tabular}}
\end{table}

\begin{figure}[htb]
\centering
\includegraphics[bb = 18 150 590 682,height=8cm,clip]{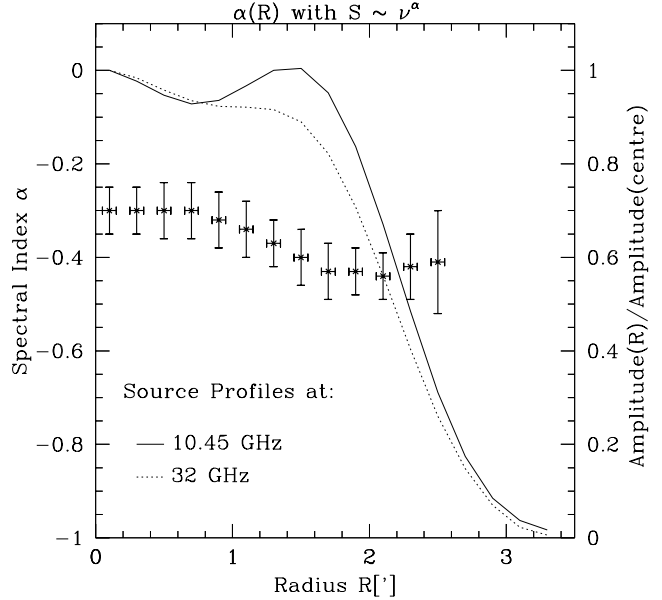}
\caption{The spectral index $\alpha$ and source profiles as a
         function of the radius R. The angular resolution is $1\farcm 15$}
   \label{alphaprof}
\end{figure}

\subsection{The radio spectrum}

The integrated spectrum of the SNR is plotted in Fig.~\ref{spectrum}. The
flux density values obtained from the Effelsberg and the Nobeyama maps
are listed in Table~\ref{flux}. Other values were taken from the
references indicated. The fitted spectral index of $\alpha = -0.50$ is
a typical value of a mature supernova remnant.

However, G11.2$-$0.3 has a flat spectrum core component.
Fig.~\ref{alphaprof} shows the radial variation of spectral index
between 10.45~GHz and 32~GHz at 1\farcm 15 angular resolution. The
spectral index between these two frequencies calculated with the
integrated flux densities from Table~\ref{flux} is $\alpha =
-0.45$, just marginally flatter than the spectral fit in
Fig.~\ref{spectrum}. The spectrum flattens from about $\alpha =
-0.45$ at the outer shell to about $\alpha = -0.3$ in the inner
part of the SNR. The $\alpha$ profile suggests a flatter radio
spectrum in the interior of the remnant. The overall spectral
index of $\alpha=-0.45$ between 10.45~GHz and 32~GHz indicates
that the flat component starts to flatten the overall spectrum of
the source.

Integrating the flux density of the compact sources gives
$110\pm20$~mJy at 1.4~GHz and $80\pm15$~mJy at 32~GHz. A spectral
index of $\alpha=-0.10\pm0.08$ results. However, there are
spectral variations. The central source is not visible at 1.4~GHz,
thus indicating an inverted spectrum. This is supported by the
observations of Green et al. (\cite{green}). At 4.76~GHz a compact
source with a peak flux density of about 2~mJy is visible which is
absent in their 1.46~GHz map at the same angular resolution. The
4.885~GHz map of Becker et al. (\cite{becker1985}) at $8\arcsec$
angular resolution barely resolves that source, but a central
enhancement of about 10 to 15~mJy/beam is seen. Of the outer
sources, the spectrum of the eastern one seems slightly steeper
than that of the western component. Clearly, sensitive higher
angular resolution data are needed to determine the spectrum of
the source components more precisely.

A separation of the shell emission and the central emission is
difficult with the present data. Based on the pulsar data we estimate
in Sect.~4.2 the flat spectrum radio emission to be about 1~Jy, which we
assume to have a spectral index of $\alpha = 0.0$. Subtracting the
plerionic component from the integrated flux density values results in
a spectral index of $\alpha = -0.57$ for the shell emission.
Fig.~\ref{spectrum} shows the resulting total spectrum when adding both
components. The differences to the spectral fit of the integrated flux
densities are quite small for the present frequency range, but would
become larger at higher frequencies.

\begin{figure*}[htb]
\centering
\includegraphics[bb = 55 185 535 695,height=16.5cm,clip]{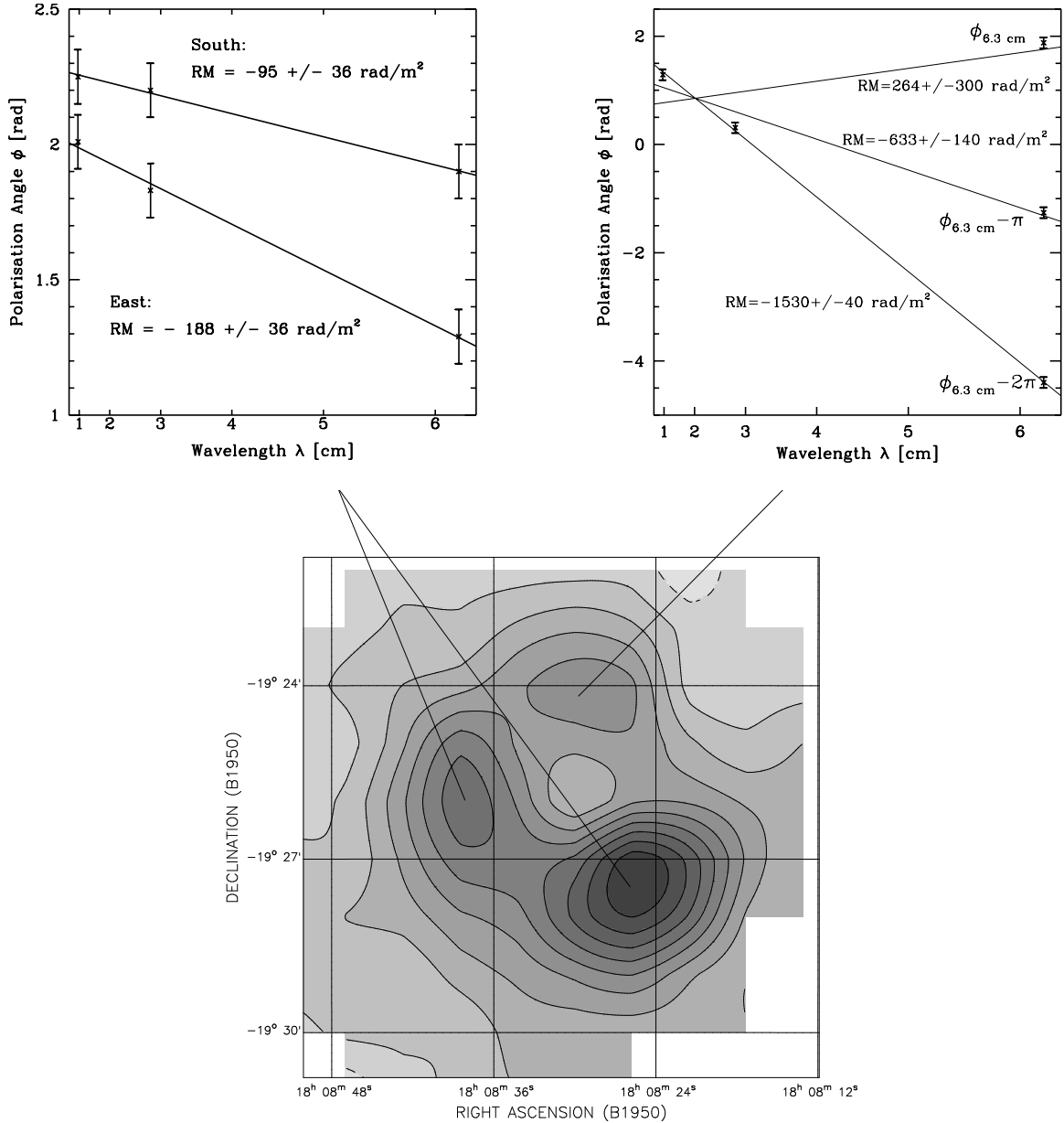}
\caption{RM values calculated between 32~GHz, 10.45~GHz, and
4.85~GHz at a resolution of 2\farcm 5. The 10.45~GHz Polarized
Intensity map
     shown is at 2\farcm 5 angular resolution}
   \label{rmlowres}
\end{figure*}

\subsection{The polarization properties}

At 32~GHz a rotation measure (RM) of 2000~rad/m$^2$ is required to
rotate the polarization angle by about 9\degr . Therefore we can
ignore Faraday rotation effects at 32~GHz. This means that the
magnetic field responsible for the synchrotron emission from the
shell of G11.2$-$0.3 has a radial orientation. The magnetic field
structure is characteristic for a young SNR in which the ejected
material, with its frozen--in radial magnetic field, dominates the
expansion and the synchrotron emission.

The integrated linearly polarized flux density (PI, see
Table~\ref{flux}) is about 2--3~\%, which is quite low for a SNR.
Nevertheless, at the three wavelengths where polarization was observed
PI is strong enough to derive rotation measures (RM). At the common
angular resolution of 2\farcm 5 we calculated RMs for three components
as marked in Fig.~\ref{rmlowres}.

\begin{figure}
\centering
\includegraphics[bb = 40 120 568 588,height=8cm,clip]{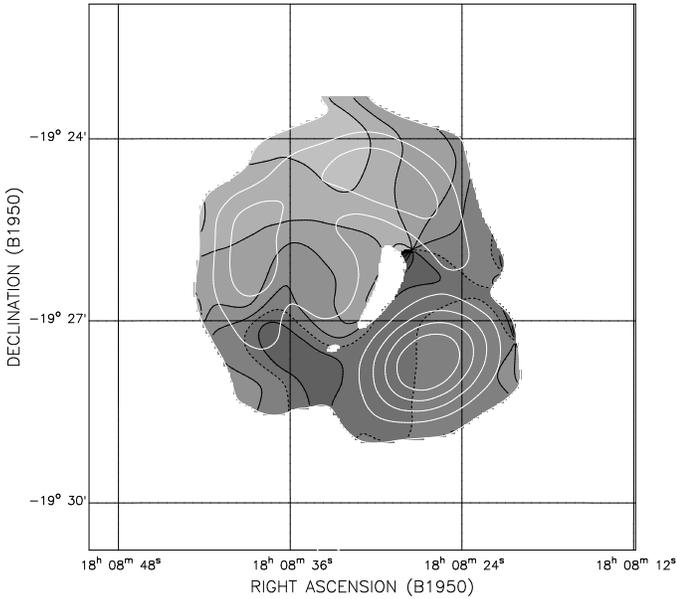}
\caption{RM map of G11.2$-$0.3, calculated between 10.45~GHz and
32~GHz at a resolution of 1\farcm 15 (greyscale and black
contours). Contours shown are at $-$1600~rad/m$^2$ to
400~rad/m$^2$ in steps of 400~rad/m$^2$. 0~rad/m$^2$ is shown
dashed. The error in RM is about 200~rad/m$^2$. White contours
represent the 10.45~GHz polarized emission at the same angular
resolution}
   \label{rmmap}
\end{figure}

\begin{figure}
\centering
\includegraphics[bb = 40 120 568 588,height=8cm,clip]{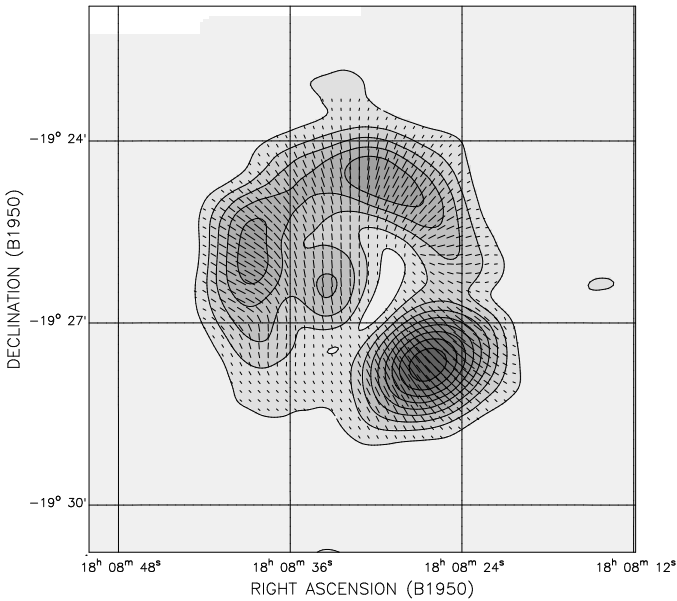}
\caption{Map of polarized intensity at 10.45~GHz with vectors in
intrinsic B-field direction calculated with the RM values from the
map shown in Fig.~\ref{rmmap}}
   \label{intrb}
\end{figure}

The southern and eastern component have low RMs, while the northern
blob shows about $-$1530~rad/m$^2$. The 4.75~GHz data suffers from
ambiguities, but not the higher frequency data. Therefore a map of the
lowest possible RM was calculated between 10.45~GHz and 32~GHz at
1\farcm 15 angular resolution. We assumed the polarization angles at
32~GHz to be correct. At 10.45~GHz we checked the original angle
$\phi$, $\phi + \pi$, and $\phi - \pi$. The angle which resulted in the
lowest RM was taken. The RM map is displayed in Fig.~\ref{rmmap}. The
PI map at 10.45~GHz shown in Fig.~\ref{intrb} is corrected for RM
effects. The SNR clearly has a radial magnetic field except for the
southeastern quadrant where the polarized intensity is very low.
Otherwise the map is fully in agreement with the 32~GHz map, which is
mostly unaffected by even high RMs. The reason for the excessive RM
in the northern part of the SNR is unknown. Either a magnetic field
enhancement or an enhanced electron density
within this part of the remnant or in its immediate surroundings
is required. The much lower RM observed towards the south and east of
the SNR rules out a significant contribution from Faraday rotation
within the interstellar medium along the line of sight between the SNR
and us.

\section{Discussion}

A spectral index of $\alpha = -0.50$ from the fit to the integrated
flux densities of G11.2$-$0.3 indicates a spectrum flatter than
all the other early phase SNRs.  Their spectral indices as listed by
Green (\cite{snrcat}) are:

{\bf
\begin{itemize}

\item Cas A: $\alpha = -0.77$

\item Kepler: $\alpha = -0.64$

\item Tycho: $\alpha = -0.61$

\item SN1006: $\alpha = -0.60$

\end{itemize}}

When taking into account the flat spectrum core the spectrum of the
shell steepens to about $\alpha \sim -0.57$. Since G11.2$-$0.3 is older
than the four listed historical SNRs the spectral difference may reflect
an evolutionary effect from free expansion to adiabatic expansion.
It is still not fully understood what exactly happens during that
transition phase and in particular under what circumstances it happens.

\begin{figure}
\centering
\includegraphics[bb = 40 120 570 680,width=8.8cm,clip]{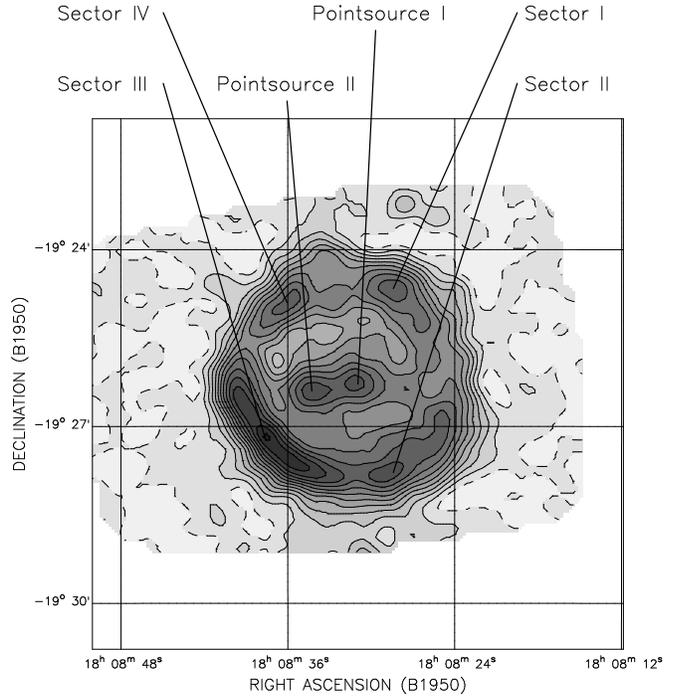}
\caption{Map of G11.2$-$0.3 in total power at 32~GHz with the same
greyscale levels as in Fig.~\ref{tpmaps}d. The different
components of the source are indicated}
   \label{defname}
\end{figure}

In the subsequent discussion we refer to six different components
of the SNR. The shell of G11.2$-$0.3 is separated into the four
prominent sectors as indicated in Fig.~\ref{defname}. Each sector
has a width of $90\degr$ starting with Sector~I from $0\degr$
(straight north) in clockwise direction. Within the SNR we refer
to unresolved sources visible at 32~GHz, which are also indicated
in Fig.~\ref{defname}. Properties of the individual components are
listed in Table~\ref{fluxsect}.

\begin{table}
\caption{Properties of the SNR's different parts
 at 32~GHz. RMs were calculated between 32~GHz and 10.45~GHz. For the
 sectors the RM values refer to the sector's
 radio continuum peak. The error for the RMs
 is about 200~rad/m$^2$.}
\begin{tabular}{lr@{$\pm$}lr@{$\pm$}lr@{$\pm$}lr}\hline
  \bf & \multicolumn{2}{l} \bf TP [mJy] & \multicolumn{2}{l}
  \bf PI [mJy] & \multicolumn{2}{l} \bf \%-Pol. &
   RM [$\frac{\rm rad}{\rm m^2}$] \\ \hline
  Sector I & 330 & 40 & 22.0 & 3.0 & 6.7 & 1.5 & $-$1150 \\
  Sector II & 420 & 50 & 27.0 & 4.0 & 6.4 & 1.5 & +130 \\
  Sector III & 380 & 40 & 3.5 & 1.0 & 0.9 & 0.5 & +550 \\
  Sector IV & 360 & 40 & 18.0 & 3.0 & 5.0 & 1.2 & $-$1430 \\
  Source I & 40 & 10 & 4.0 & 2.0 & 10.0 & 6.0 & $-$900 \\
  Source II & 40 & 10 & 4.0 & 2.0 & 10.0 & 6.0 & $-$750 \\ \hline
\end{tabular}
    \label{fluxsect}
\end{table}

\subsection{The structure of the shell}

All the shell sectors look very similar in total intensity. As
already mentioned by Green et al. (\cite{green}), the structure of
the shell appears rather clumpy without a sharp outer boundary. An
exception is Sector~III, which has a higher peak flux and is well
defined by an outer boundary in contrast to the others (see
Fig.~\ref{tpmaps}). This indicates a later phase of evolution for
this part of the SNR. This assumption is supported by the very low
percentage polarization in Sector~III, while all the other sectors
show polarization between 5 and 7 \% (Table~\ref{fluxsect}).

In early stages of SNR evolution most of the expanding material is
ejecta from the supernova explosion itself. A radial magnetic field
is frozen into this material. In the free expansion phase
the ejecta with the radial magnetic field dominate the hydrodynamics
and the synchrotron emission. Since a radial magnetic field is observed
in the shell of G11.2$-$0.3 we conclude that the SNR is in this
phase of evolution. In the later development more ambient material is
swept up by the expanding shock wave. This material is compressed and
collected within a thin shell behind the shock wave, outside the ejecta.
The ambient magnetic field component frozen into this shell is
perpendicular to the expansion direction which means
the expansion direction which means perpendicular to the magnetic field
in the ejecta. The percentage
polarization will decrease until the contribution of the swept up
magnetic field to the synchrotron emission equals that of the ejecta.
At this point we expect no linear polarization unless the observations
are able to resolve these structures, which is not the case here.
Sector~III seems to be exactly at that stage while the other sectors
are still in an earlier stage of evolution.

This advanced evolutionary phase of Sector~III is also implied by
the X--ray observations. The soft X--ray luminosity of a SNR
depends on the amount of emitting mass which is of course
increasing with time while the expanding shock wave sweeps up
interstellar matter. The soft X--ray luminosity also depends on
the emissivity of this material in the soft X--ray band which
mainly depends on its temperature. In early phases the temperature
exceeds $10^8$~K by a factor of five or even more. Such a high
plasma temperature peaks in hard X--rays. During the free
expansion phase the soft X--ray luminosity increases with
decreasing temperature and an increase of emitting material. In
the adiabatic expansion phase the temperature changes are rather
small. Thus the soft X--ray luminosity increases just depending on
the increasing mass in the expanding shell. In the free expansion
phase the soft X--ray luminosity increases much faster than in
later evolutionary phases of the SNR. This gives a very good
indication of evolutionary differences within the remnant.

Sector~III shows the strongest emission in the ROSAT HRI image
(see Fig.~\ref{xrayradio}). This implies that Sector~III is the
most developed part of the SNR. Apparently the `aging' of the
remnant seems to increase from Sector~I to Sector~III, which is
likely due to different densities of the ambient medium. The
required density increase points away from the Galactic Plane.
This is either a local effect or the density peak at about 5~kpc
distance is below $b = -0.3\degr$. However, no such density
gradient is visible in the distribution of \ion{H}{i} from the
channel maps of the Effelsberg \ion{H}{i}-survey by Braunsfurth \&
Rohlfs (\cite{brau}) at about 9\arcmin\  angular resolution.
Therefore any density contrast has to be small. This assumption is
supported by the nearly circular shape of the SNR, while a density
variation would lead to different expansion velocities.

X--ray observations indicate that between 3 and 4~M$_\odot$ of
interstellar material has been swept--up so far (Vasisht et al.\
\cite{vasisht}; Reynolds et al.\ \cite{reynolds}). A tangential
magnetic field is frozen into this material, which interacts with
the ejecta by forming a convective zone between both layers. In this
zone material from both layers and their magnetic field lines are
mixed by turbulence. Synchrotron emission is expected from all these
zones. Since linear polarization with a radial magnetic field is
observed, the mass of the ejecta should be considerably larger than the
mass of the swept up material. Only a type II supernova
explosion will eject such a large amount of material. The pulsar found
within the remnant supports this conclusion. There is no indication
for a stellar wind bubble around the SNR from the \ion{H}{i} data
(Braunsfurth \& Rohlfs\ \cite{brau}). This suggests an early B-type
progenitor star with a mass of about 10--15~M$_\odot$. These stars have no
high velocity winds or high mass loss rates. A supernova explosion of a B0 or
a B1 star would fit all observations.

\subsection{The interior structures}

Torii et al. (\cite{torii1997}) detected an X--ray pulsar in the
geometrical center of G11.2$-$0.3 within the hard spectrum
X--ray synchrotron nebula previously discovered by Vasisht
et al. (\cite{vasisht}). The exact position of these components and
therefore their physical relation to the central radio sources cannot
be determined very precisely. The positional error is about $\pm
1\arcmin$, which makes it difficult to identify the synchrotron nebula
of the pulsar with either of the three radio components within the
remnant. However, it is most likely that the pulsar coincides with the
central component visible in the ROSAT-HRI image (Reynolds et al.\
\cite{reynolds}). In Fig.~\ref{xrayradio} a greyscale representation of
this observation is shown together with white contours indicating the
32~GHz measurement in total power. Evidently the most likely radio
counterpart of the hard X--ray synchrotron nebula is the central source
(Source I) with an inverted spectrum. The flux density listed in
Table~\ref{fluxsect} is, however, confused with a contribution from the
western component showing up at 1.4~GHz (Fig.~\ref{interior}).

With the correlation between a pulsar's rotational energy loss rate
$\dot{\rm E}$, the radio surface brightness $\Sigma$ and the diameter
D of its synchrotron nebula found by Kothes (1998) we can estimate the
flux density of this nebula in G11.2$-$0.3. With $\dot{\rm E} = 8.8\cdot
10^{36}$~erg/s and a mean diameter of $1\arcmin$ at a distance of 5~kpc
we get a flux density of about 1~Jy at 1~GHz. Assuming a spectral index of
$\alpha_{\rm plerion} = 0.0$ for this nebula we get $\alpha_{\rm shell}
= -0.57$ (see Fig.~\ref{spectrum}) for the shell.

The total flux density of the central component is much larger
than the integrated flux density of the compact sources
(Sect.~3.2.). Most of the flat spectrum emission is therefore
diffuse. The compact polarization feature seen at 32~GHz
(Fig.~\ref{pimaps}) is slightly offset to the north of the compact
structures and therefore related to the diffuse emission. Its RM
is about $-$800~[rad/$\rm m^{2}$], rather similar to that of the
compact sources I and II. The central RMs follow the gradient of
RM across G11.2$-$0.3 as seen in Fig.~\ref{rmmap}.

We note some similarities of G11.2$-$0.3 with the crab--like SNR
G21.5$-$0.9, where a compact X--ray source (Slane et al.\
\cite{slane}), which has no radio counterpart at 22~GHz, is seen
at the center of axisymmetric radio structures suggesting some
collimated outflow of particles (F\"urst et al.\  \cite{fuerst}).
Certainly the angular resolution of the present data does not
allow a decision as to whether this scenario also holds for
G11.2$-$0.3. The possibility of confusion of the central emission
with extragalactic background sources can be ruled out as the high
resolution maps by Green et al. (\cite{green}) and Becker et al.
(\cite{becker1985}) resolve the eastern and western sources of the
NVSS map into diffuse extended emission.

\section{Conclusions}

Our high frequency observations confirm the historical SNR
G11.2$-$0.3 as a composite SNR. The structure of the centrally
peaked flat spectrum component, which is powered by electron and
magnetic field injection by the central pulsar, is partly
resolved. At 32~GHz we have an indication of a central source with
an inverted spectrum, which likely coincides with the pulsar. The
two outer symmetrical components were previously resolved at high
angular resolution and might result from a collimated outflow.
However, this needs to be confirmed by more sensitive data. A
strong gradient in RM across G11.2$-$0.3 is observed with highest
absolute RM values in the northern part of the SNR. The SNR shows
a radial magnetic field over most of the shell structure. The
shell was found to be in the transition phase from free to
adiabatic expansion, with differences between the four sectors of
the shell. This suggests that the interacting interstellar
material likely has different densities. Combining available
X--ray and radio data we conclude that G11.2$-$0.3 is likely the
remnant of a type II supernova explosion with an early type B
progenitor star.

\begin{acknowledgements}
RK acknowledges the support from Prof. Richard Wielebinski and the
MPG through a PhD and a post-doctoral fellowship at the MPIfR,
when most of the observations were made. This research made use of
data obtained through the High Energy Astrophysics Science Archive
Research Center Online Service, provided by the NASA/Goddard Space
Flight Center. We wish to thank Ernst F\"urst, Lloyd Higgs, and
Tom Landecker for critical reading of the paper. We are grateful
to the referee John Dickel for helpful comments.
\end{acknowledgements}

\end{document}